\documentclass[a4paper, 12pt]{article}
\usepackage[utf8]{inputenc}
\usepackage[english]{babel}
\usepackage{amsmath, amssymb, amsfonts}
\usepackage{fullpage, indentfirst, cite}
\usepackage{epsfig, graphicx, graphics}
\usepackage{color}

\newcommand{\be}{\begin{equation}}
\newcommand{\ee}{\end{equation}}

\title{
{\Large Hadronically decaying heavy dark matter and high--energy neutrino limits}
\date{}
\author{M.~Yu.~Kuznetsov\footnote{mkuzn@inr.ac.ru}
\vspace{.2cm}\\
\footnotesize \it Institute for Nuclear Research of the Russian Academy of Sciences, \\
\footnotesize \it 60th October Anniversary Prospect 7a, 117312 Moscow, Russia}
}

\begin{document}

\begin{flushright}
INR-TH-2016-042
\end{flushright}

{\let\newpage\relax\maketitle}

\begin{abstract}
We consider dark matter consisting of long--living particles with masses \\
$10^{7}~\lesssim~M~\lesssim~10^{16}$~GeV decaying through hadronic
channel as a source of high energy neutrino.
Using recent data on high energy neutrino from IceCube and Pierre Auger experiments we
derive the upper-limits on neutrino flux from dark matter decay and
constraints on dark matter parameter space. For the dark matter masses
of order $10^8$ GeV the constraints derived are slightly stronger than
those obtained for the same dark matter model using the high energy gamma-ray limits.
\end{abstract}


\section{Introduction}
The idea that dark matter consists of heavy long--living particles
was proposed in the context of inflationary cosmology.
There are several mechanisms of production of these
particles that are able to yield the observed
relic abundance. Among them are production in non-equilibrium plasma,
production during the decay of inflaton (preheating) and production by non-stationary
gravitational field~\cite{Zeldovich:1971mw, Zeldovich:1977, Kuzmin:1997jua, Berezinsky:1997hy,
Kofman:1994rk, Khlebnikov:1996zt, Khlebnikov:1996wr, Kuzmin:1998kk, Chung:1998rq, Chung:1998zb, Kuzmin:1998uv}.
Although, heavy dark matter was also discussed
irrespectively of inflation~\cite{Khlopov:1987bh, Fargion:1995xs, Gondolo:1991rn}.
It was also realised that heavy decaying particles can be the source of ultra high energy cosmic rays (UHECRs)
that evade the GZK cutoff~\cite{Kuzmin:1997jua, Berezinsky:1997hy}. Although, the absence of the GZK cutoff is not confirmed
by the modern cosmic--ray experiments~\cite{AbuZayyad:2012ru, Abraham:2008ru} the heavy dark matter is still
under consideration as a possible source of high energy cosmic rays,
in particular photons and neutrino.

The heavy dark matter candidate $X$ has two main parameters: mass
$M_X$ and lifetime $\tau$. The case of absolutely stable $X$--particles
is not so interesting from the experimental point of
view --- its annihilation cross--section is bounded by unitarity:
$\sigma_X^{\rm ann.} \sim 1/M_{X}^2$, making their indirect detection
impossible for the today experiments~\cite{Gorbunov:2011zz}.
The direct detection of these particles, whether stable or long--living,
is also experimentally unreachable due to their small number density.
However, there are several sources of constraints for the heavy dark--matter parameters: the mass is subject
to cosmological constraints~\cite{Kolb:1998ki, Kuzmin:1998kk, Kuzmin:1999zk, Chung:1998zb, Chung:2004nh, Gorbunov:2012ij},
and the lifetime of the dark--matter particles can be effectively constrained
using the observed fluxes of various high energy particles or limit on these fluxes.
For example, in Ref.~\cite{Kalashev:2008dh} the constraints have been put using the shape of charged
cosmic--ray spectra. However, with the modern cosmic ray data this method
bounds $\tau$ not so well as gamma--ray and neutrino flux limits.
Various gamma--ray data and limits was employed to constrain heavy dark--matter parameters
in Refs.~\cite{Murase:2012xs, Cohen:2016uyg, Aloisio:2015lva, Esmaili:2015xpa, Kalashev:2016cre}.

The detection of the high energy neutrino events by IceCube experiment~\cite{Aartsen:2013jdh, Aartsen:2014gkd}
has attracted significant attention. There were many works interpreting these events as
an astrophysical neutrino signal~\cite{Kalashev:2014vra, Aartsen:2015rwa, Dermer:2014vaa} as
well as a dark matter decay signal~\cite{Murase:2015gea, Bhattacharya:2014vwa, Esmaili:2013gha, Dev:2016qbd, Esmaili:2014rma, Cohen:2016uyg}.
At the same time, the constraints on various models of neutrino origin have been placed~\cite{Rott:2014kfa}.
There were also pre-IceCube studies where neutrino limits were employed to constrain
heavy dark matter parameters~\cite{Esmaili:2012us, Murase:2012xs}.
This study is mainly inspired by the publication of the new refined sample of the IceCube high--energy
neutrino data along with the updated exposure of this experiment~\cite{Aartsen:2016ngq}.
In that work the stringent cuts were employed to eliminate the atmospheric neutrino background.
The resulting data set contains only two events with PeV order energy, both consistent with the
astrophysical neutrino Monte--Carlo. This fact together with the non-observation of higher
energy events allows the IceCube collaboration to place limits on the astrophysical neutrino flux
and to constrain several models of astrophysical neutrino origin.

In this work we use the same data sample to place limits on the neutrino flux
from the decay of dark matter with masses $10^{7} \lesssim M_X \lesssim 10^{16}$ GeV
and to constrain its lifetime. For comparison we also derive constrains using
Pierre Auger Observatory data~\cite{Aab:2015kma} that reports non-detection
of neutrino with energies $E_\nu \gtrsim 10^{17}$ eV.
This study complements our previous research~\cite{Kalashev:2016cre},
where heavy decaying dark matter parameters was constrained
by the high energy gamma--ray limits.

\section{Neutrino flux from dark matter decay}
\label{flux}
In this study we consider dark matter consisting of heavy scalars $X$
decaying through the channel $X \rightarrow q \bar{q} \rightarrow \nu_i \; (\bar{\nu_i}).$
We assume that all quark flavors are coupled to $X$ similarly.
The decay through this channel can be described
irrespectively of the particular form of $X$--quarks coupling,
since the most important physical phenomenon of relevance is hadronisation,
see Refs.~\cite{Aloisio:2003xj, Sarkar:2001se} for the details of this approach.
It should be noted that other possible decay modes,
e.g. those related to gauge bosons, may also lead to comparable neutrino flux,
however we do not consider theses modes in the present study.
Some results and constraints related to heavy dark matter
decaying into neutrino via various channels can be found in
Refs.~\cite{Murase:2012xs, Esmaili:2012us, Esmaili:2015xpa, Cirelli:2010xx}.
The main difference between the present study and these works is
that we consider the DGLAP evolution of the fragmentation functions (see below)
that allows us to handle the wider range of the dark matter masses.

The method of calculation
of the final state stable particles spectra for the hadronic decays of heavy particles
was reviewed in our previous work~\cite{Kalashev:2016cre} and mainly follows the 
Refs.~\cite{Aloisio:2003xj, Sarkar:2001se}. In this study we consider the neutrino 
flux. The main contribution to the flux comes from the decay of charged pions via processes
\be
\label{decays}
\pi \rightarrow \mu \nu_\mu, \quad
\mu \rightarrow e \nu_\mu \nu_e\, .
\ee
There are also contributions from kaons as well as from charmed mesons but they
are an order of magnitude smaller than the contribution of pions.
Moreover, the uncertainty of the pion flux which is dominated
by the uncertainty of the pion fragmentation functions on
the initial energy scale is of the same order
as the contributions of other mesons to the neutrino flux~\cite{Hirai:2007cx}.
Therefore we assume that the neutrino production is saturated by the pion decays.
Using the results of Ref.~\cite{Cirelli:2010xx} we are also make sure of negligibility of
electro--weak corrections to the decay spectrum.

We consider the spectrum of pions $\frac{dN_{\pi}}{dx}$, where $x=\frac{2 E_\pi}{M_X}$,
produced in the decay of $M_X$. It can be obtained by the evolution of the
pion fragmentation functions from the initial scale to the $M_X$ scale via DGLAP equations~\cite{GLD, AP}:
\be
\frac{\partial D_i^\pi(x,s)}{\partial \ln s} = \sum_j \frac{\alpha_s(s)}{2\pi}P_{ij}(x,\alpha_s(s)) \otimes
D_j^\pi(x,s)\,,
\ee
where $D_i^\pi(x,s)$ is the fragmentation function of pion from the parton $i$, $s$ is the factorization scale,
$\otimes$ denotes the convolution $f(x) \otimes g(x) \equiv \int_x^1 dz/z f(z)g(x/z)=
\int_x^1 dz/z f(x/z)g(z)$ and $P_{ij}(x,s)$ is the splitting function for the parton branching $i \rightarrow j$.
We use the same assumptions about DGLAP evolution and fragmentation functions as
in our previous work~\cite{Kalashev:2016cre}, namely we assume that all quark flavors are coupled to gluon similarly
and consider the mixing of gluon fragmentation function with the quark singlet fragmentation function.
As in our previous work we use the code of Ref.~\cite{Aloisio:2003xj} to solve DGLAP equations
numerically in the leading order of $\alpha(s)$. We take the initial fragmentation
functions parametrized on the scale of $1$ GeV from Ref.~\cite{Hirai:2007cx} and extrapolate
them to the interval $10^{-5} \le x \le 1$.

\begin{figure}
   \includegraphics[width=13.50cm]{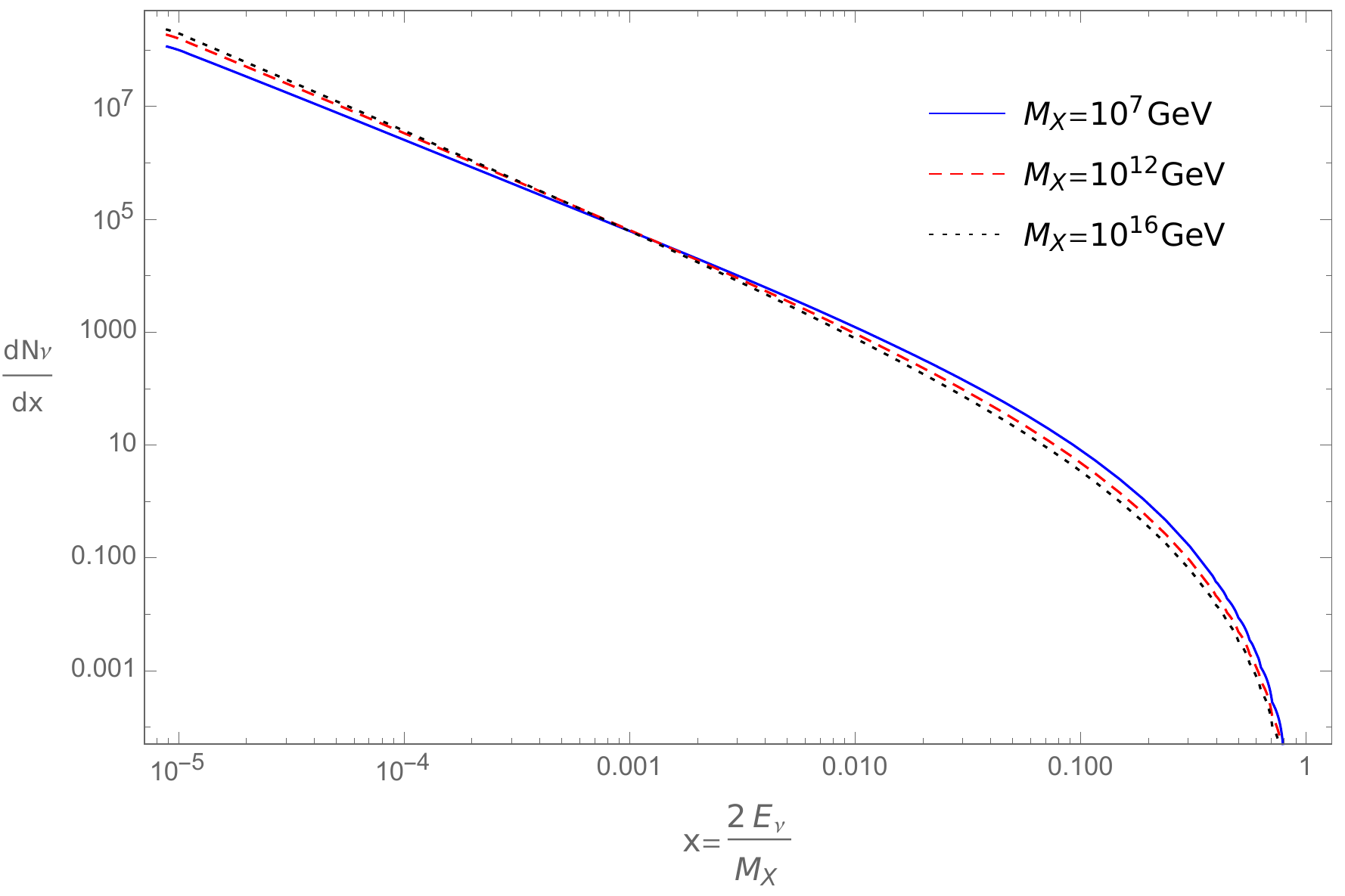}
   \caption{The total spectra of neutrino and antineutrino from $X$ particle
    decay for three different values of $M_X$.}
   \label{injection_spectra}
\end{figure}

The neutrino spectrum from pions decay is given by
\be
\frac{dN_{\pi \rightarrow \nu_\mu}}{dx} = 2\: R \int\limits_{x R}^1 \frac{dy}{y} \: \frac{dN_{\pi}}{dy} \,,
\ee
while the neutrino spectrum from the decay of secondary muons is 
\be
\frac{dN_{\mu \rightarrow \nu_i}}{dx} = 2 \int\limits_x^1 \frac{dz}{z} f_{\nu_i}\left(\frac yz \right)  \:
\frac{dN_{\pi}}{dz} \,;
\ee
where $r = (m_\mu/m_\pi)^2 \simeq 0.573$, $R=\frac{1}{1-r}$ and
the functions $f_{\nu_i}(x)$ are taken from Ref.~\cite{Kelner:2006tc}:
$$
f_{\nu_i}(x)=g_{\nu_i}(x)\,\Theta(x-r)+(h^{(1)}_{\nu_i}(x)+h^{(2)}_{\nu_i}(x))\,\Theta(r-x)\,,
$$
$$
g_{\nu_\mu}(x)=\frac{3-2r}{9(1-r)^2}\,\left(9x^2-6\ln x-4x^3-5\right),
$$
$$
h^{(1)}_{\nu_\mu}(x)=\frac{3-2r}{9(1-r)^2}\,\left(9r^2-6\ln r-4r^3-5\right),
$$
$$
h^{(2)}_{\nu_\mu}(x)=\frac{(1+2r)(r-x)}{9r^2}\left[9(r+x)-4(r^2+rx+x^2)\right] \ ,
$$
$$
g_{\nu_e}(x)=\frac{2}{3(1-r)^2}\,\left[(1-x)\,\left(6(1-x)^2 + r(5+5x-4x^2)\right) + 6r\ln x \right],
$$
$$
h^{(1)}_{\nu_e}(x)=\frac{2}{3(1-r)^2}\,\left[(1-r)\,\left(6-7r+11r^2-4r^3\right) + 6r\ln r \right],
$$
$$
h^{(2)}_{\nu_e}(x)=\frac{2(r-x)}{3r^2}\left(7r^2-4r^3+7xr-4xr^2-2x^2-4x^2r\right) \, .
$$
The examples of neutrino spectra from the decay of $X$ particles
with different masses are shown in Fig.~\ref{injection_spectra}.

Neutrinos propagate cosmological distances unattenuated. The resulting flux that reaches
the Earth consists of the galactic and extragalactic parts.
The initial flavor composition of the pion decay products is modified by
the neutrino oscillations during the propagation. We assume the flux reaching the Earth
is completely mixed, i.e. the flavor ratio $\nu_e : \nu_\mu : \nu_\tau$ is $1 : 1 : 1$.
We also assume that neutrinos are radiated isotropically in the decay of $X$ particle.
For the galactic neutrino flux calculation we use the Navarro--Frenk--White dark matter
distribution~\cite{Navarro:1995iw, Navarro:1996gj} with the parametrization for the Milky Way 
from  Ref.~\cite{Cirelli:2010xx}. Being strongly anisotropic, the galactic signal
has to be convolved with the exposure of the particular experiment to obtain the
perceived flux (see next Section). Contrary, the extragalactic flux is isotropic
and undergoes the cosmological redshifting
\be
\frac{dN^{\rm EG}_\nu}{dE_\nu}\left(E_\nu\right) = \frac{1}{4\pi M_X \tau} \int\limits_0^\infty
\frac{ \rho_0 \, c/H_0 }{\sqrt{\Omega_m (1+z)^3 + (1-\Omega_m)}}\frac{dN_\nu}{dE_\nu}\left(E'_\nu\right) dz
\ee
where $c/H_0 = 1.37 \cdot 10^{28}$ cm is the Hubble length, $\rho_0 = 1.15 \cdot 10^{-6} \; {\rm GeV}/{\rm cm}^3$ is the average
cosmological dark matter density for today, $\Omega_m = 0.27$ and the injected spectrum $\frac{dN_\nu}{dE_\nu}$
is taken as a function of neutrino energy at redshift $z$: $E'_\nu = E_\nu (1+z)$.

\section{Analysis \& discussion}
\label{results}
\begin{figure}
   \includegraphics[width=13.50cm]{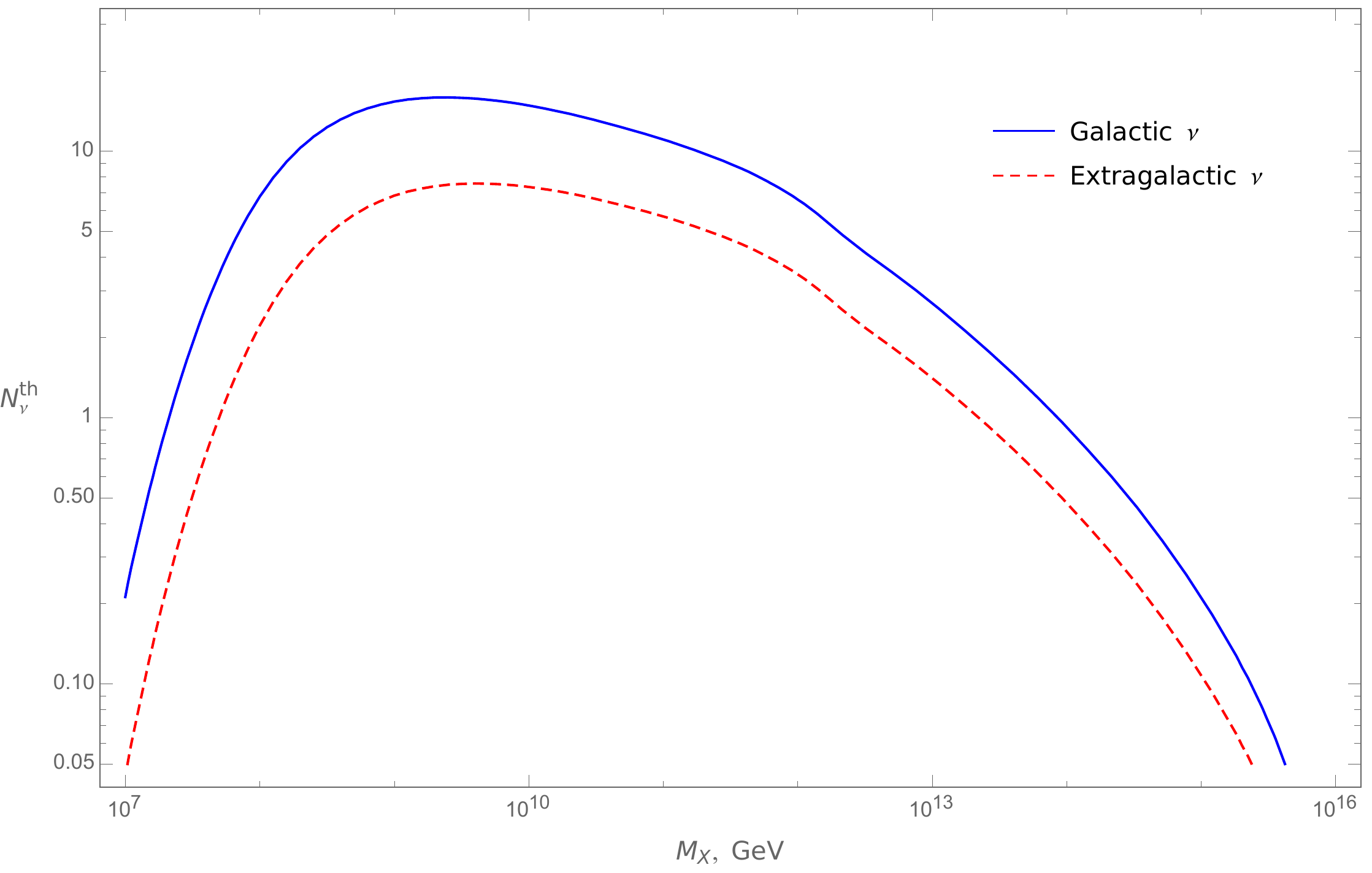}
   \caption{Total number of galactic (solid line) and extragalactic (dashed line) neutrinos from decays
   of dark matter particles with various masses $M_X$ and lifetime $\tau = 10^{20}$
   yr as it could be received by the IceCube experiment.}
   \label{N_G-N_EG}
\end{figure}

The method of constraining the dark--matter parameters with neutrino limits slightly differs
from that using with the gamma--ray limits. The exposure of neutrino observatory
depends on the neutrino energy, therefore flux limits depend on neutrino spectrum.
Below we briefly describe the method.
The quantity one needs to compare with the observation is the total number of neutrino
events that would be detected in the given experiment under the assumption of the given neutrino spectrum.
The method of calculation of this quantity was described in Ref.~\cite{Anchordoqui:2002vb}.
Below all the quantities are related to neutrino therefore the index $\nu$ is omitted.
For the galactic neutrino flux one has
\be
N_{\rm G} = \frac{1}{4\pi M_X \tau} \int\limits_{\Delta E} \int\limits_{V} \rho\left[R(r,\delta,\alpha)\right]\:
\varepsilon(E, \delta, \alpha)\: \frac{dN}{dE}(E)\: \cos(\delta)\: dr\: d\delta\: d\alpha\: dE\; ;
\ee
where $\rho[R]$ is a dark matter density as a function of distance from the Galactic Center $R$,
$r$ is a distance from Earth, $\varepsilon$ is the exposure of the given
observatory as a function of the neutrino energy $E$ and equatorial coordinates $\{\delta, \alpha\}$.
The integration takes over all volume of the dark--matter halo ($R < 260$ kpc)
and over the entire range $\Delta E$ of the neutrino energies accessible for a given observatory.
In practice, the exposure is given for several bands of zenith angle, averaged over each band.
For IceCube we adopt the exposure as a function of declination
(which uniquely translates to zenith angle in the case of IceCube) and energy as it is
given in Ref.\cite{Abbasi:2011ji} and normalize it to the actual IceCube exposure
of Ref.~\cite{Aartsen:2016ngq}. For Pierre Auger we use the exposure given in Ref.~\cite{Aab:2015kma}
together with the formula of the effective exposure of extensive air shower observatory~\cite{Sommers:2000us, Aab:2014ila}:
\be
\omega(a_0,\delta,\theta_\text{max}) \sim (\cos a_0\,\cos\delta\,\sin\alpha_m+\alpha_m\sin a_0\,\sin\delta),
\ee
where $a_0$ is the geographical latitude of the given observatory, $\theta_\text{max}$ is the maximal zenith
angle accessible for fully efficient observation in this observatory and $\alpha_m$ is given by
\be
\alpha_m=\begin{cases}
0 & ;\xi>1,\\
\pi & ;\xi<-1,\\
\arccos\xi & ; -1 < \xi < 1\,;
\end{cases}
\ee
\be
\xi = \frac{(\cos\theta_\text{max}-\sin a_0\,\sin\delta)}{\cos a_0\,\cos\delta}.
\ee
The number of events from the extragalactic flux is
\be
N_{\rm EG} = \int\limits_{\Delta E} \varepsilon(E)\: \frac{dN_{\rm EG}}{dE}(E)\: dE\; ;
\ee
where the exposure $\varepsilon(E)$ is integrated over the celestial sphere.
Thus the total number of events predicted by the theory is
\be
\label{N_total}
N_{\rm th} = N_{\rm G} + N_{\rm EG}\, .
\ee
The example of $N_{\rm G}$ and $N_{\rm EG}$ for fixed $\tau$ and various masses $M_X$ is shown
in Fig.~\ref{N_G-N_EG}.
There are two factors of resulting neutrino signal enhancement. One is
due to the observation of the galactic flux, which exceeds the contribution
of the rest of the Universe as one can learn from Fig.~\ref{N_G-N_EG}. Another one
is due to the fact that largest high--energy neutrino observatories --- IceCube and Pierre Auger
can observe the enhanced neutrino flux from the Galactic Center region which
is located in the southern sky.

For each mass $M_X$ the lifetime $\tau$ is subject to constrain.
The standard technique of Ref.~\cite{Feldman:1997qc} implies that we vary $\tau$
until the predicted number of events $N_{\rm th}$ reaches from below the number $N_{\rm limit}$ specified for
a given number of observed events $N_{\rm obs}$, number of background events $N_{\rm bg}$ and given confidence level. 
We may calculate $N_{\rm th}$ over full range of accessible energies or in separate energy intervals.
In the latter case the constraints on parameter $\tau$ can be weaker,
since the number $N_{\rm limit}$ does not depend on the length of the energy interval.
In the case when $N_{\rm bg}=0$ and $N_{\rm obs}\ne 0$, the other method is more
appropriate. We split the full energy range in separate intervals $\Delta E^i$ with certain
$N_{\rm obs}^i$ in each one and generate Monte--Carlo set which places in the $i$-th interval
the number of events $N_{\rm MC}^i$ following the Poisson distribution with the mean
$\lambda^i = N_{\rm th}^i$, the theoretical number of events calculated in
the respective energy interval $\Delta E^i$. For each particular value of the parameter
$\tau$ we generate a large number of these Monte--Carlo realisations.
Then we vary the parameter $\tau$ until the fraction of realisations
where $N_{\rm MC}^i > N_{\rm obs}^i$ at least in one bin reaches the given confidence level C.L.
In the case of all $N_{\rm obs}^i=0$ this method yields
the same results as the Feldman--Cousins technique.
While for $N_{\rm obs} > 0$ the constraints of the Monte--Carlo method appears somewhat
stronger. In the IceCube dataset we neglect the background of $0.064^{+0.023}_{-0.039}$ atmospheric
neutrino events and therefore can apply the described method.

\begin{figure}
   \includegraphics[width=13.50cm]{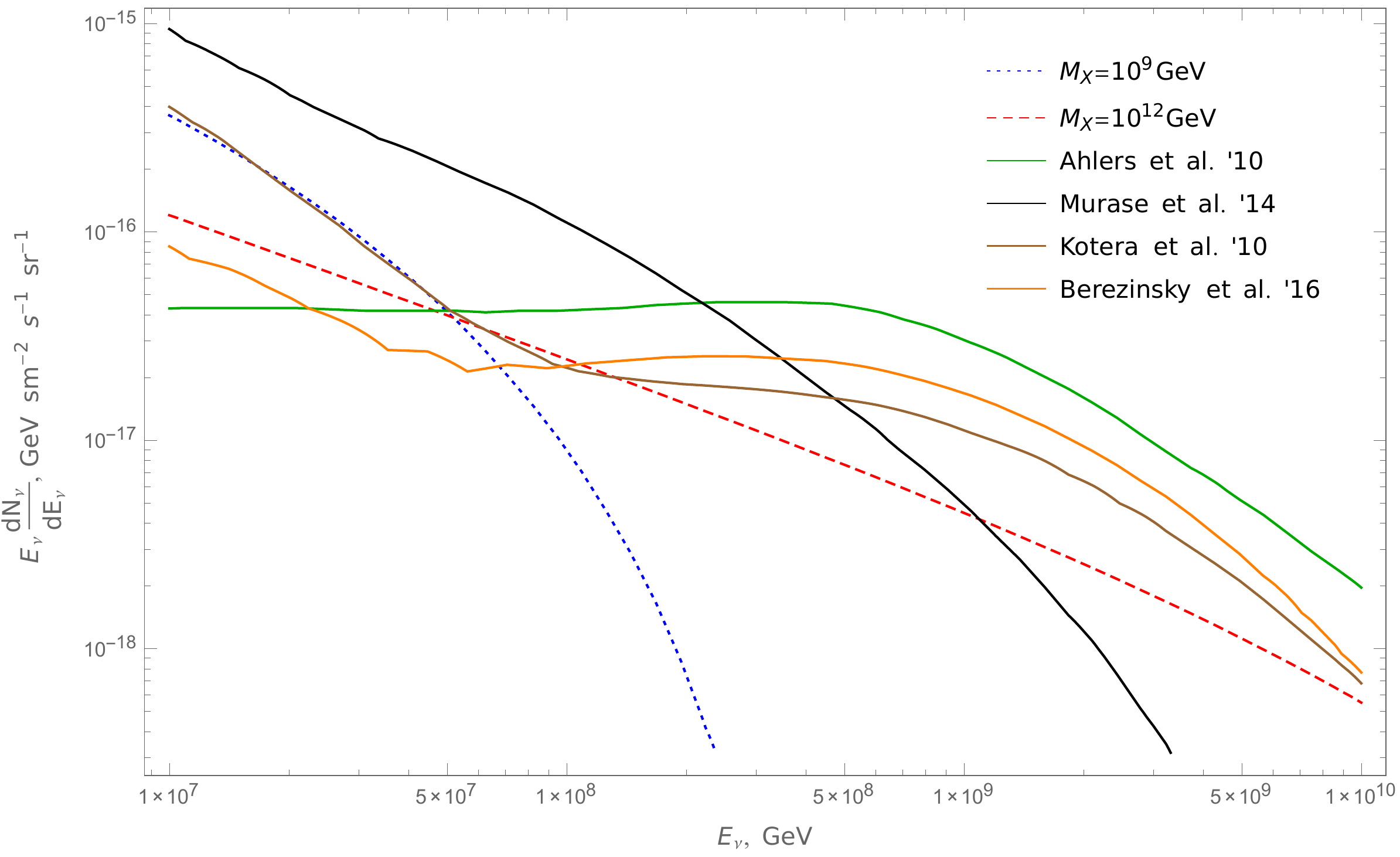}
   \caption{All--sky averaged neutrino fluxes from decays of dark--matter particles with masses $M_X=10^{9}$ and $M_X=10^{12}$ GeV
    and marginally allowed lifetime ($\tau= 6.6 \cdot 10^{20}$ and $\tau= 3.5 \cdot 10^{20}$ yr respectively)
    compared with various models of astrophysical~\cite{Murase:2014foa} (solid black) and
    cosmogenic~\cite{Berezinsky:2016jys, Kotera:2010yn, Ahlers:2010fw} neutrino fluxes (the sum of neutrino and
    antineutrino of all flavours).}
   \label{fluxes1}
\end{figure}

\begin{figure}
   \includegraphics[width=13.50cm]{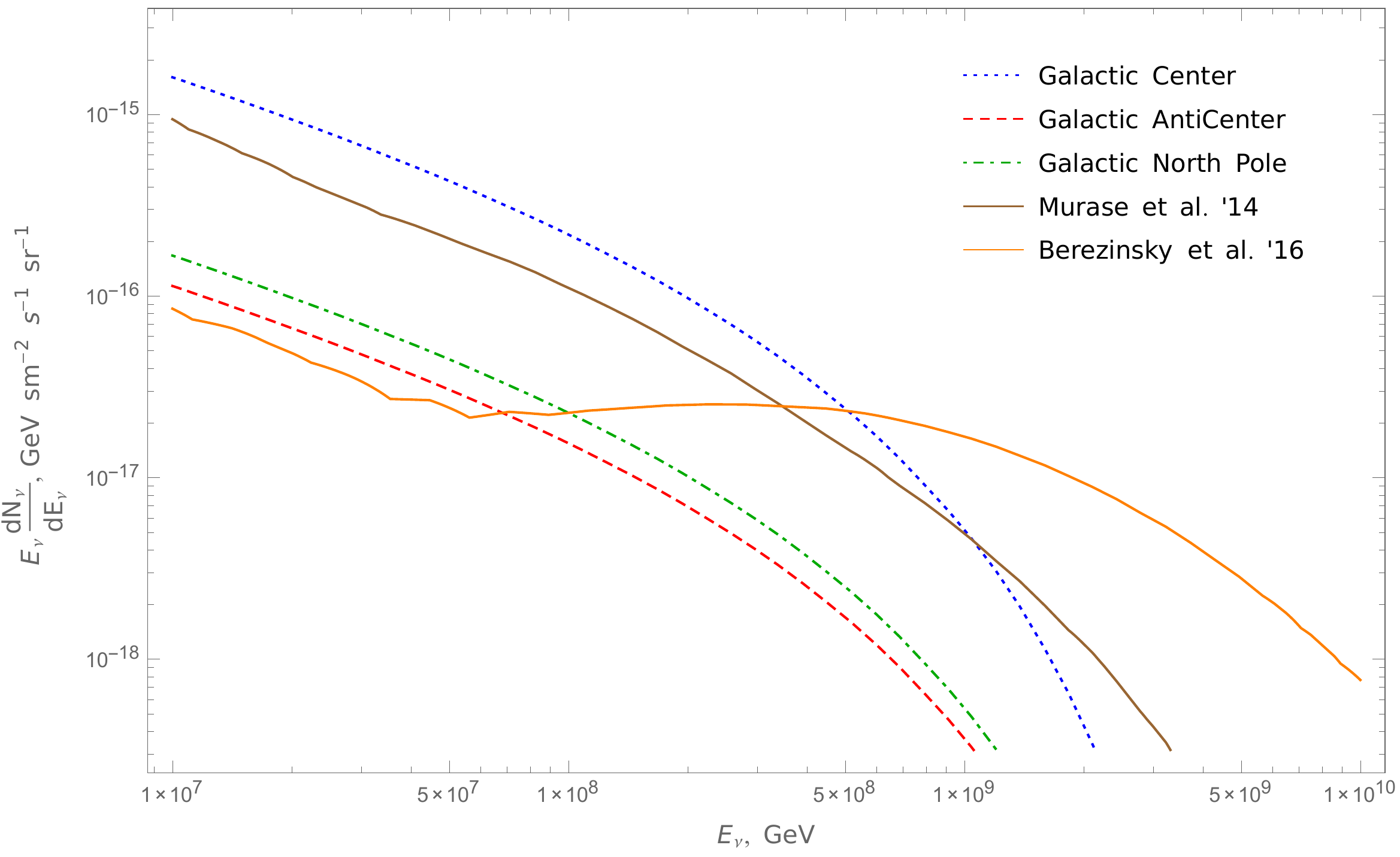}
   \caption{Comparison of neutrino fluxes from decays of dark--matter particles with mass $M_X=10^{10}$ GeV
    and marginally allowed lifetime $\tau= 7.75 \cdot 10^{20}$ yr, coming from several directions
    with one model of astrophysical neutrino flux~\cite{Murase:2014foa} (solid brown) and one model
    of cosmogenic one~\cite{Berezinsky:2016jys} (solid orange) (the sum of neutrino and
    antineutrino of all flavours).}
   \label{fluxes2}
\end{figure}

\begin{figure}
   \includegraphics[width=13.50cm]{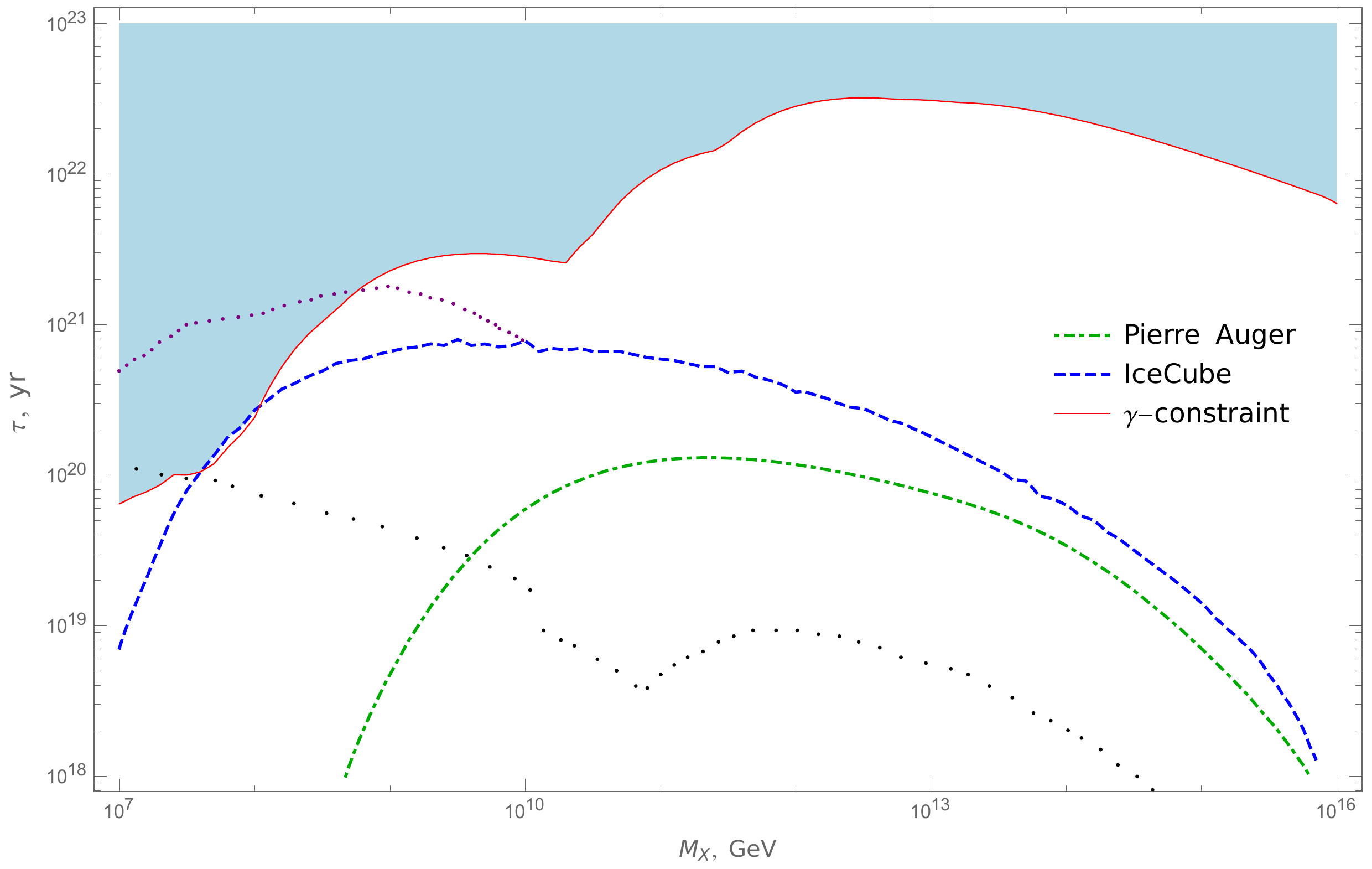}
   \caption{$90\%$ C.L. exclusion plot for mass $M_X$ and lifetime $\tau$ of dark--matter particles. White area is excluded.
   For comparison we present the constraints obtained with photon limits~\cite{Kalashev:2016cre} (solid thin red line). We also
   show the constraint obtained in the dark matter model with $X \rightarrow \nu\bar{\nu}$ decay channel~\cite{Esmaili:2012us} (black dots)
   and constraint for $X \rightarrow b\bar{b}$ channel which assumes that the IceCube events are of astrophysical
   origin~\cite{Cohen:2016uyg} (purple dots).}
   \label{exclusion_plot}
\end{figure}

The constraints on the parameter space $\{M_X, \tau\}$ are presented in Fig.~\ref{exclusion_plot} together
with the constraints of works~\cite{Esmaili:2012us, Cohen:2016uyg} as well as the
gamma--ray constraints obtained in our previous work~\cite{Kalashev:2016cre}.
We should note that the present constraints are conservative since we consider the total
predicted neutrino flux as a product of the dark--matter decay and do not allow for the possible
astrophysical or cosmogenic contribution.
One can see that the gamma--ray constraints overlap the neutrino ones in almost all dark--matter mass range
except the narrow region around $M_X \sim 10^8$ GeV, where the neutrino constraints is slightly stronger.
Nevertheless neutrino observation remains a crucial tool for the dark--matter indirect detection.
For example, in the model of hadronically decaying dark matter considered in this paper
and in our previous work~\cite{Kalashev:2016cre}
the ratio of neutrino flux to photon flux have the certain value $r$ which variates in the range
$0.8 \lesssim r \lesssim 1.8$ depending on energy and $M_X$. This ratio could be an
additional criterion for distinguishing between various hypotheses of photon and neutrino fluxes origin.

Some examples of neutrino fluxes from the decay of the dark matter with the marginally allowed
lifetime are shown in Figs.~\ref{fluxes1}---\ref{fluxes2} together with some competing astrophysical
and cosmogenic neutrino fluxes.
In Fig.~\ref{fluxes1} the all--sky averaged fluxes are given, while
in Fig.~\ref{fluxes2} we show the fluxes coming from the several directions related to our Galaxy.
One can see that it is hard to distinguish the all--sky averaged fluxes of dark--matter decay
from astrophysical and cosmogenic ones. However, this problem simplifies when we compare the
directional fluxes~\footnote{There is a subtlety related to the choice of dark matter profile.
Navarro--Frenk--White profile produces the larger signal from the Galactic Center direction compared to the cored profiles.
For example, the GC flux in Burkert profile~\cite{Burkert:1995yz} is approximately 3 times smaller than that of
NFW, while the difference between the overall fluxes is negligible.}.
Therefore the analysis of the Galactic anisotropy of the signal
become crucial for the dark matter indirect search.

\section{Conclusion}
The implications of the new IceCube dataset of high energy neutrino to
the hadronically decaying heavy dark matter theory was considered. It was found
that for the dark matter masses $10^{7} \le M_X \le 10^{16}$ GeV
the neutrino data bound dark matter lifetime stronger than the gamma-ray
limits of the extensive air shower observatories only in the narrow region around
$M_X \sim 10^8$ GeV. One of the reasons of this fact
is that photon exposures of experiments are typically larger than its neutrino exposures.
It is also meaningful that the non-zero flux of high energy neutrino was observed,
contrary to the non-observation of photons of the same energies.
It was emphasized that the relevant test for distinguishing the signal
of the decaying dark matter from signals of other origin
is the analysis of the galactic anisotropy and photon--neutrino flux ratio. 

\section*{Acknowledgements}
I would like to thank S.~Troitsky, G.~Rubtsov, O.~Kalashev and D.~Gorbunov for helpful discussions.
I am especially indebted to R.~Aloisio, V.~Berezinsky and M.~Kachelriess for providing
the numerical code solving the DGLAP equations. This work has been supported
by the Russian Science Foundation grant 14-12-01340.

\suppressfloats

\end{document}